# Visible evidence to magnetism of graphene oxide

Ling Sun [a,b]

[a] *Beijing Guyue New Materials Research Institute, Beijing University of Technology, 100 Pingleyuan, Chaoyang District, Beijing 100124, China.*
[b] *Advanced (Energy) Materials Joint Laboratory, Beijing University of Technology, 100 Pingleyuan, Chaoyang District, Beijing 100124, China.*
Correspondence E-mail: sunling@bjut.edu.cn

**ABSTRACT**

Graphene oxide continues to amaze scientific community for multiple potentials in a broad span of applications such as catalysts, adsorbents, oxidants, etc., determined by constant unveiling of its fantastic natures. Of them, magnetism is not ultimately identified and directly observed by naked eyes. Herein, we report graphene oxide directionally migrated and deposited together simply under external magnetic field from common Nd-Fe-B magnet, whereas the ferromagnetism of graphene oxide did not exhibit that strong as iron. Therefore, we illustrated this interesting pathway to keep close to such 2D carbon material and potentially promoted magnetic-oriented applications.

**Keywords**
Magnetism, Graphene oxide, Direct evidence

# 1. Introduction

Graphene oxide (GO) is one oxygenated derivative of graphene, which is accessible by means of facilely oxidizing graphite or expanded graphite. [1-3] We also regard GO as one dimensional macromolecular matter due to the massive content of carbon as well as surface implanted oxygen-containing functionalities, such as carboxyl, epoxy, carbonyl and hydroxyl, etc. All these groups are found to some extent specifically distributed on the graphene plain, e.g., edge sites for carboxyl. A decade year ago, it was reported that the chemical composition of the edges produces ferromagnetism to graphene, one of that-mentioned impurity-free carbon materials.[4] While in 2010, physically exfoliated graphene was reported only a little paramagnetic, but strong diamagnetic like graphite.[5] In the same year, a paper from theoretical calculation demonstrated the unpaired spin induced by epoxy group rendered GO of magnetism and also to control the magnetism with heat or chemical treatment.[6] However, no direct visible evidence has been ever reported to demonstrate magnetism scenarios in graphene nanostructures, about which a review published in 2010 complaint as we thought. Even so far, this scenario has not been changed a bit. Herein, we intend to show our observation from a simple experiment. We prepared GO from our previously modified Hummers method. To construct magnetic field, commercial Nd-Fe-B magnets were used. The observation was consequently achieved by putting diluted GO aqueous solution across the magnets and keeping them stand still over hundred hours.

# 2. Materials and methods

## 1.1. materials

Commercial expanded graphite was purchased from Qingdao Huatai Lubricant Sealing S&T Co. Ltd., Qingdao, China. Sulfuric acid, potassium permanganate, and other chemicals were from Sinopharm Chemical Reagent Co., Lt. unless specifically noted, and all were used as received without further purification. Sintered neodymium-iron-boron rectangular shape magnets (Nd-Fe-B N35) were purchased online from www.taobao.com, with a dimension 100 mm×50mm×20mm.

## 1.2. Preparation of graphene oxide

Graphene oxide preparation was made following to our previous report. In that report, we devised some modifications to the Hummers method using industrially size-limited expanded graphite. A typical reaction was conducted as follows. Expanded graphite (5 g) and potassium permanganate (15 g) were first mixing simply by hand in a bottom-round flask (500 mL) with ice-water bath. The mixture temperature was controlled beneath 10 ℃ as a safety measure. Then, 150 ml of concentrated sulfuric acid was added to the mixture with continuously mixing, by hand, or mechanical arm. Subsequently a uniform liquid paste was obtained. Then the mixing was withdrawn. The whole paste was controlled beneath 25-28 ℃ by a ice bath. (Once the temperature intended to decrease, the water bath was moved away). The liquid paste inside the flask was put still at ambient condition to continue as-called mid-temperature reaction (~27 ℃) until a foam-like intermediate formed (~2 hrs), featuring a large volumetric expansion. After that, deionized water (400-500 ml) was added, and the whole flask was kept at the temperature no lower than 85 ℃ for at least 1.5 hrs. A dark brown suspension turned out along with mechanical stirring. As the reaction proceeded 2 more hours, the above suspension turned thicker and became yellowish brown in color and later gradually turned darkening. The suspension was filtered and subjected to repeated dialyses in deionized water with the purpose to remove impurities, such as $H^+$, $SO_4^{2-}$, $Mn^{2+}$, etc. Typically, we tested the pH and remnant $SO_4^{2-}$ of water by using pH test stripes (>6) and 0.1 M barium chloride solution (no white precipitate), respectively. Before storage for further application, the above suspension was given a further centrifugation at 9,000 rpm, 0.5 h per cycle. After that, kinds of GO solutions were then obtained.

## 3. Results and discussion

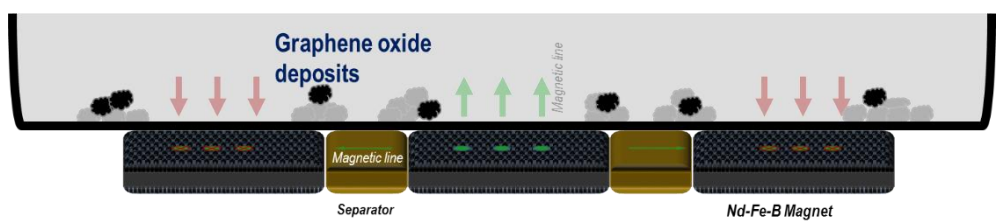

**Figure 1. Diagram for testing GO-magnetic field interaction**

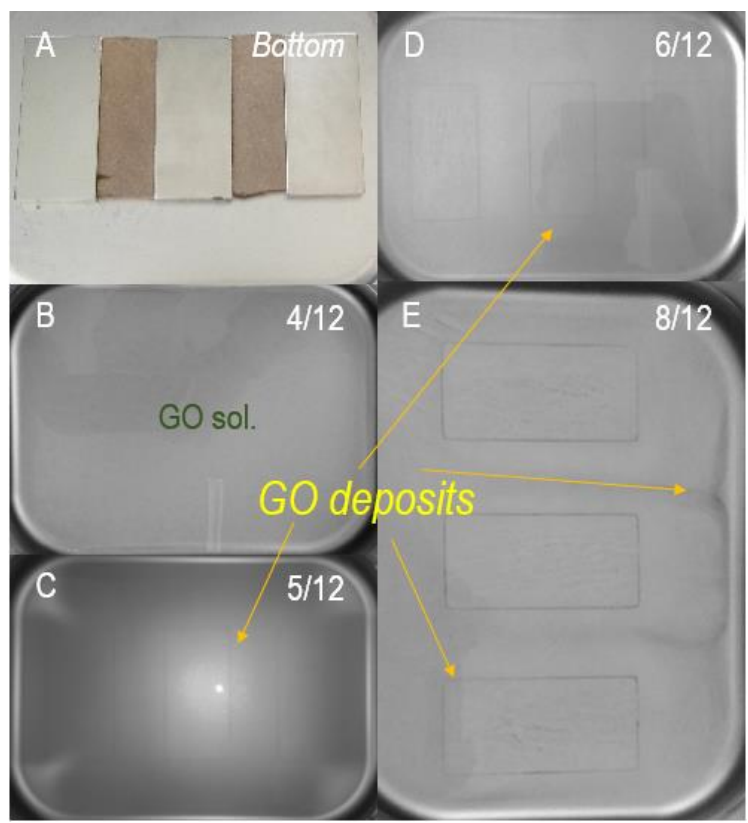

**Figure 2. Time-evolvement of GO solution in Nd-Fe-B magnetic field**

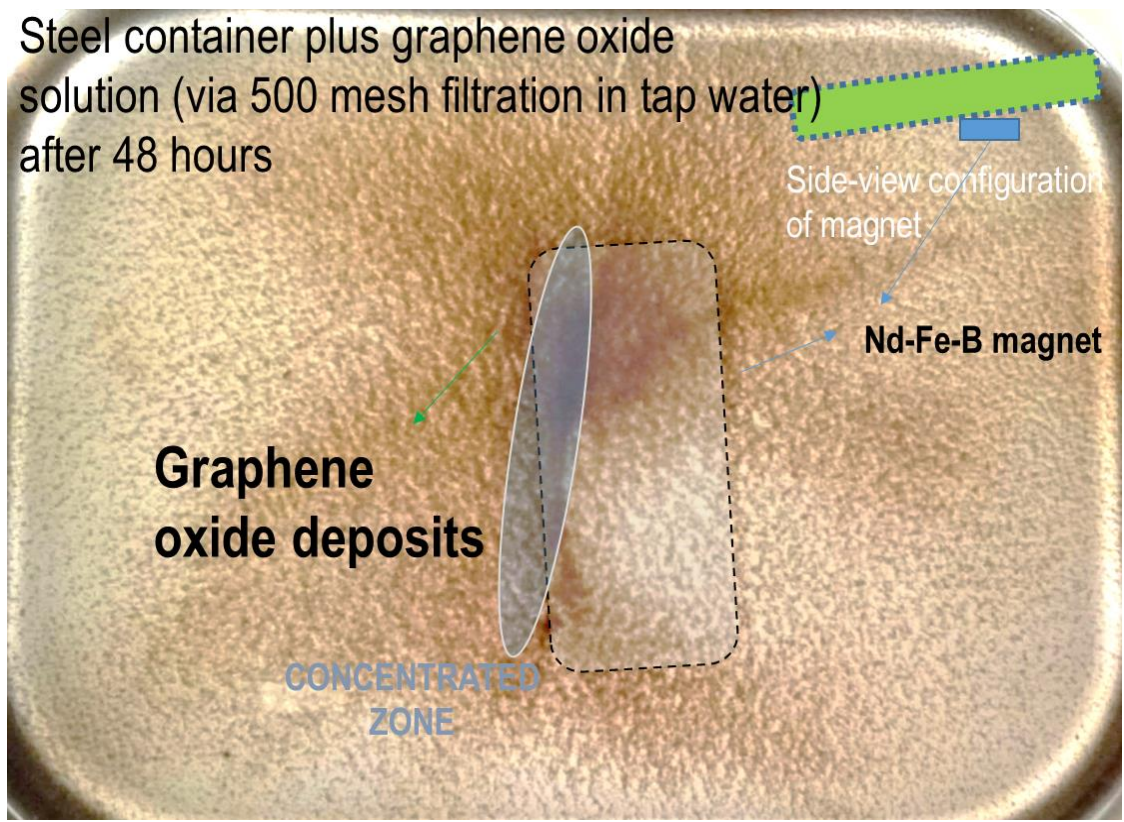

**Figure 3. Magnetism expression of GO in tap water**

GO (~0.15%) was prepared via Hummers method and characterized by a series of conventional characterizations to ascertain the reactants structures and morphologies (*data not shown*). After that, we collected GO specifically in the upper-layer of a long-term standing GO solution. As we know, resultant graphene oxide sheets from graphite are usually with different dimensions, graphene layer numbers. After standing for hours or days, GO solution would delaminate layer by layer. That is, much smaller molecular-weight and thinner GO sheets are the most major solute in upper layer, as compared to those distributing or depositing near/in the bottom. To better observe the magnetization behavior of GO, atom thick GO guaranteed was chosen as the most basic requirement.

About 15-ml GO were dripped and further diluted in deionized water in a bottom-flat washbasin, with the whole taking on well transparence and a light yellow color. Notably, without shaking, this dispersing process demonstrated rather slow, while suffering a simple shake, GO dispersed fast and seconds late, to become homogeneously distributed inside water phase.

We designed this experiment as partly schemed in Figure 1. A liquid-storing vessel is laid over the magnets with strong magnetism. The targeted GO is in the vessel while the magnetic field comes from a battle of magnets very nearby. The interaction between the targets and magnetic field would ultimately results in any positive/negative conclusions, which in fact depends on natural property of GO here. Later, we realized such setup first, through simply sticking a row of magnets on the washbasin bottom (Figure 2A). And we further checked the magnet field whether it was not heavily resisted by material of washbasin, simply tested by one more magnet to approach the upper face of basin bottom. Actually we indeed felt very strong magnetic force there. Then the aforementioned GO solution was added. This experimental then proceeded into observation. In Figure 2B-E, we show a time dependence over the liquid-phase change. It is interesting to find differences occurred day by day. In detail, in the first day, the solution was initially homogeneous without deposits (Figure 2B). Through 24 hours, three rectangular shapes appeared in the bottom of the washbasin, which showed identical shapes to these magnets of the other side. Definitely, GO sheets were comprised into these deposits. However, as mentioned above, specifically-selected GO was used to be highly water dispersible and stable, if without external disturbance to storage circumstances, such as heating, chemically redox, etc. The deposits were conclusively a definite result from the magnetic interactions over GO sheets. Noticeably, even we roiled the solution once more, there again came with the same result. Thus, as-synthesized GO sheets demonstrated steady magnetic property, being capable of interacting with the magnetic field.

As mentioned in the head section, several groups had theoretically calculated or microscopically measured magnetic property of graphene [5, 7] (no ferromagnetism, weak paramagnetism, graphite-like strong diamagnetism), graphene oxide (ferromagnetism at ground state of zigzag graphene nanoribbon [6]) and even graphite oxide [8] (ferromagnetism). Most magnetism were discovered to have relationship with the properties of edge structures and specific functional groups (for example, epoxy group [6]). Similar to the other magnetic stuffs, the GO was in priority gathered in these high-magnetic-field producing zones, such as the edges, tips, which also schemed in Figure 1 and consequently verified in Figure 2 and Figure 3 (in tap water, rich in some additional salt ions). Detailed information, especial the effect of supportive substrate or additional ionic strength on the expression of GO magnetism as well as condense states, is expected to going on further investigation. Thus, this work relative to this experiment is still undergoing.

## 4. Conclusions

GO indeed delivered a comprehensive magnetism, and we observed them regularly migrating in a magnetic field from commercial Nd-Fe-B magnet. This finding inspired us of developing more interesting applications with GO.

**Acknowledgement**

The author is greatly thankful to Mr. Sun YB, Mrs. Fei XD, Dr. Chen WY, and Miss Sun TY for their living and spiritual support. The work was partially supported by the financial support from Beijing University of Technology (No.105000514116002,105000546317502).